\documentclass{PoS}

\title{Recent progress on light scalars: from confusion to precision using dispersion theory}

\ShortTitle{Recent progress on light scalars: from confusion to precision. }

\author{\speaker{J. R. Pel\'aez}\\
        Departamento de F\'{\i}sica Te\'orica II. Facultad de CC. F\'{\i}sicas. Universidad Complutense. 28040 Madrid. SPAIN\\
        E-mail: \email{jrpelaez@fis.ucm.es}}

\abstract{In this talk I briefly review the recent developments on light scalar meson spectroscopy, 
paying particular attention to the causes of major revision of the $\sigma$ or $f_0(500)$ meson in the Review of Particle Phsycis. This resonance, despite playing a central role in the nucleon-nucleon attraction as well as the QCD chiral symmetry breaking, has suffered a longstanding controversy which has been acknowledged to be finally settled. The combination of new and precise data together with rigorous dispersive approaches has turned the old confusing situation about the properties of these mesons, and even thir existence in some cases, into a field which now aims at precision studies.}

\FullConference{Xth Quark Confinement and the Hadron Spectrum,\\
		October 8-12, 2012\\
		TUM Campus Garching, Munich, Germany}

\begin{document}

\section{Introduction}

For researchers outside the field, it may come as a surprise that, despite having established 40 years ago that Quantum Chromodynamics (QCD) is the theory governing the Strong Interaction, its lowest mass spectrum, particularly that of mesons, may be still under debate. 
Actually, light scalar mesons have been a longstanding puzzle in our understanding of strong interactions, although they are very relevant both for Nuclear and Particle Physics. For the former because they are largely responsible for nucleon-nucleon attraction, and for the latter due to their role in spontaneous chiral symmetry breaking and the identification of glueballs --- two fundamental features of QCD. 

For people working outside the Hadron Physics community, this relatively poor understanding is expected from
the theory side, since it is textbook knowledge that QCD becomes non-perturbative at low energies and does not allow for a precise calculation of the spectrum, requiring non-perturbative and complicated lattice calculations. However, I have found that the younger ``outsiders''
are very surprised by the fact that the empirical properties and even the existence of many of the lightest mesons and resonances are still actively discussed, although many of them were proposed several decades before the advent of QCD. Concerning the older non-practitioners, since the situation on how many states exist, what are their masses, etc... has remained rather confusing for many decades, they tend to think that no rigorous conclusion and no progress can be made about light scalars. Admittedly, the way that, for instance, the lightest meson---the $\sigma$ resonance---has been listed in the Review of Particle Physics (RPP) \cite{PDG}, which for long has considered it a well established state while simultaneously quoting a mass between 400 and 1200 MeV..., has not helped a lot in conveying the rigorous efforts that were pursued both by theoreticians and experimentalists within the community. Fortunately, there has been a major improvement in the last RPP 2012 edition \cite{PDG}, at least for $\sigma$ particle; perhaps the most controversial light meson for many years.  

The purpose of this brief review talk is to make an account of the recent developments since the previous ``Quark Confinement and the hadron Spectrum Conference'' held in 2010. 
With few exceptions---mostly older dispersive analyses or studies using analyticity---, I will therefore concentrate on references that appeared after or at the end of 2010, which, of course, does not mean that there are other, previous, works of relevance in the field.
Thus, for the stated purpose I will follow two paths. First, the most conservative and consensual one, based on the new additions and changes in the RPP, whose tables are used by the Particle Physics community as the basic reference for particle properties. Second, my own personal view, which is less conservative, but probably closer to the one held by the majority of the community working nowadays on light scalars. As a matter of fact, the major changes in the latest RPP edition had already been widely accepted by most practitioners for more than a decade, although it is only now that these developments have made it to the RPP. I will review how the changes in the RPP, particularly that of the $\sigma$ or $f_0(500)$, have been triggered not only by the newest data, but by the existence and consistency of several rigorous and model independent dispersive approaches. These rigorous analysis do not only exist for the $\sigma$, but for other light scalars as well, most notably the $K_0(800)$, and I expect these developments should lead to further revisions in the RPP tables within the near future.

Hence, in order to illustrate the previous situation and the present state of the art, I will be referring, for simplicity and also due to the limited space, not only to the latest 2012 RPP \cite{PDG} edition, but to previous ones as well. In particular, the 2012 RPP lists the $K_0^*(800)$ as ``needs confirmation'', the $f_0(1370)$ is listed with a huge mass range, from 1200 to 1500 MeV,  it includes an $f_0(1200-1600)$ under ``further states'', and a relatively similar situation is found for vector mesons above 1 GeV. Of particular relevance for our purposes is  the RPP ``Note on light scalars below 2 GeV''. In brief, this note shows that the main caveats to these particles come from the use of conflicting data sets and model-dependent analysis, leading to huge systematic uncertainties and, I may add, very often to unphysical artifacts. However, it has been possible to overcome these caveats by combining
rigorous and model independent approaches with new data, and provide very convincing proof of the existence and properties of these states, whose latest developments I will review next.

I will spend most of the space discussing on the $\sigma$ and the major 
change it has suffered in the RPP, but I will also comment on the other scalars, like the  $f_0(980)$, $a_0(980)$, whose existence and properties are less controversial,  as well as the
$K_0^*(800)$, which is still a subject of debate and, although widely accepted by the ``light meson practitioners'', is still classified as ``not-well established'' by the RPP 2012 edition.
There are other scalars with the quantum numbers of the $\sigma$ and the $f_0(980)$, but above 1 GeV: the $f_0(1370)$, $f_0(1500)$ and $f_0(1710)$, etc....  Although affected by 
similar problems with systematic uncertainties, conflicting data sets, and use of model dependent approaches that also affect the $\sigma$, for this talk I will consider these states as heavy and discuss only those below 1 GeV. Actually, this is part of an additional heated controversy about light scalars, which is their nature problem of their classification in multiplets. In particular there are rather strong arguments for the  assignment of the $f_0(500)$, $K^*_0(800)$, $f_0(980)$ and $a_0(980)$ to the same light scalar nonet, but this debate lies beyond the
scope of this review and thus I just refer to some relevant references \cite{nonet,otherssigmakappa}
as well as to the ``Note on scalar mesons below 2 GeV'' in the RPP, and references therein.

\section{The $\sigma$ or $f_0(500)$ meson. A major change in the RPP. }

In order to gain perspective on the significance of the latest RPP major revision about the $\sigma$ meson, let me provide a historical sketch of $\sigma$ appearances, by no means complete, but enough to illustrate the confusing situation of light scalars over the last decades. 
A relatively light scalar-isoscalar field, i.e. with zero isospin, was postulated nearly 60 years ago \cite{Johnson:1955zz} in order to explain the nucleon attraction, and was soon incorporated into simple models of the Strong Interactions, like the Linear Sigma Model \cite{GellMann:1960np}, from which it gets its common name: the $\sigma$ resonance, although nowadays it is called the $f_0(500)$. In this model, the $\sigma$ is the massive remainder of a multiplet of scalars that suffers an spontaneous symmetry breaking and, with the exception of the $\sigma$ field, all become Goldstone bosons.  This realization of chiral symmetry is rather simple due to the linear realization of its symmetries. Not only in this model, but on general grounds, the $\sigma$, which has the quantum numbers of the vacuum,  is expected to play a very important role in the dynamics of the QCD spontaneous symmetry breaking. Let us also remark that the $\sigma$ also has the expected quantum numbers of the lightest glueball, which is 
one of the most remarkable features of a confining non-abelian gauge theory like QCD. Clarifying the existence and properties of the $\sigma$ is thus important for our understanding of the nucleon-nucleon attraction, the spontaneous chiral symmetry breaking of QCD, and the identification of glueballs. 

\begin{figure}[htbp]
  \centering
  \includegraphics[width=0.8\textwidth]{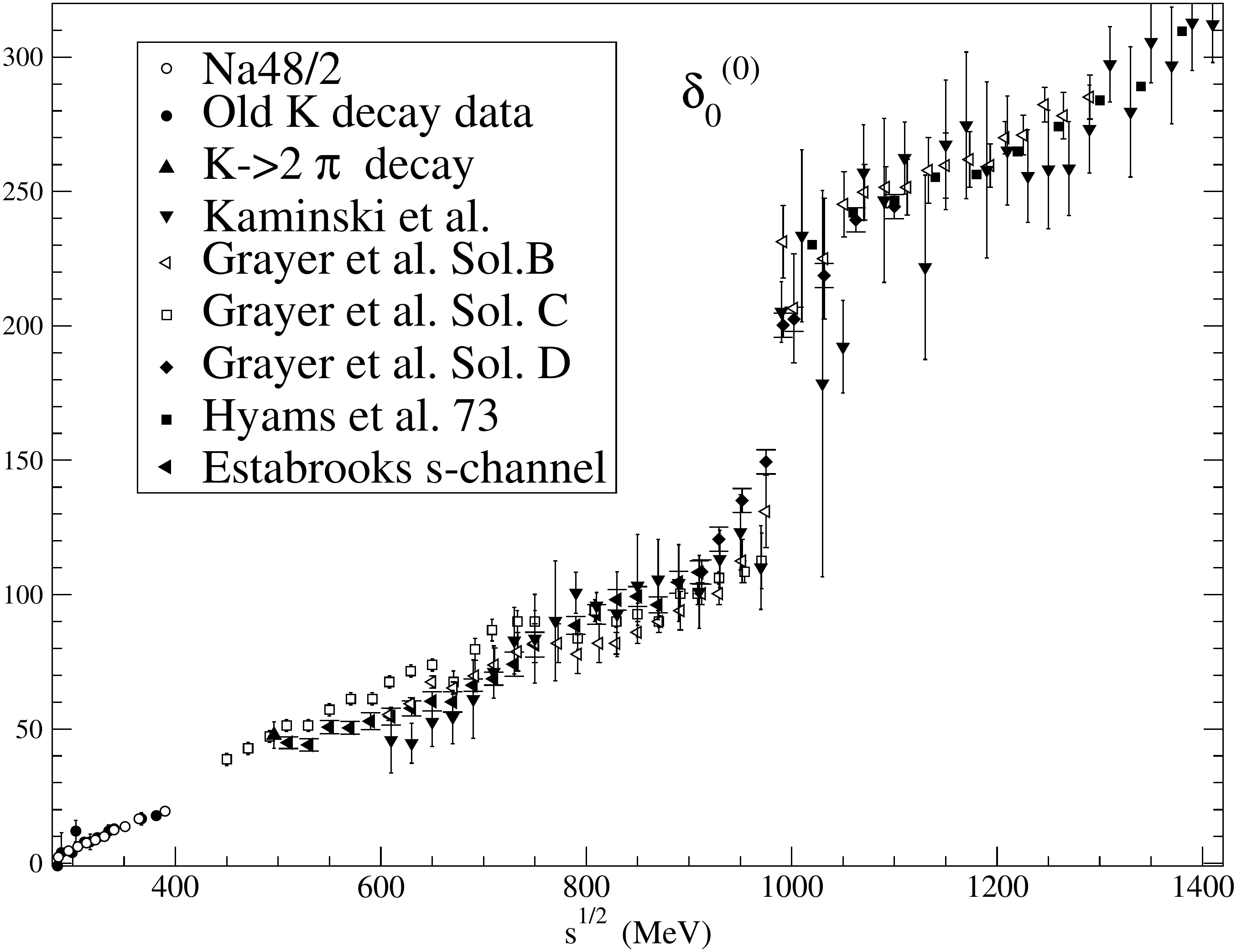}
\vspace*{.1cm}
  \caption{ Data on the scalar-isoscalar $\delta_0^{(0)}$ phase shift of $\pi\pi\rightarrow\pi\pi$ scattering\cite{Pr73,Cern-Munich,Estabrooks:1974vu,Kaminski:1996da}. Note the large differences even within data sets coming from the same experiment \cite{Cern-Munich} due to systematic uncertainties. The precise and consistent data sets below 400 MeV all come from $K\rightarrow\pi\pi\ell\nu$ decays \cite{Rosselet:1976pu,Batley:2010zza}. Also, note that there is no Breit-Wigner shape around 500-600 MeV. A Breit-Wigner shape over a background phase of about 100 degrees is seen around 980 MeV, corresponding to the $f_0(980)$.
}
  \label{fig:00data}
\end{figure}

However, and despite its relevance, from the first RPP edition until 1974, the $\sigma$ meson was considered in the RPP as a ``not-well established'' state, disappeared for 20 years after 1976, returning in 1996 under the name of $f_0(600)$, but was only declared ``well established'' in 2002, although with a surprisingly huge mass uncertainty ranging from 400 to 1200 MeV and a similarly large range, from 500 to 1000 MeV, for the width. There are several reasons for this confusing coming in and going out of the tables. First, the nucleon-nucleon potential is intuitively understood in terms of the exchange of bosons in a t-channel, i.e., not produced directly as a resonance in the s-channel. Thus, this interaction is not very sensitive to the details of the particles exchanged, even less so if they are very wide, as it is the case of the $\sigma$. Hence, traditionally, many of the lightest mesons have been studied in meson-meson scattering, where this resonances can be produced in the s-channel, particularly in $\pi\pi\rightarrow\pi\pi$, or in systems were meson-meson scattering is needed as a part of a larger process. Unfortunately, $\pi\pi\rightarrow\pi\pi$ scattering has to be extracted from $\pi N\rightarrow \pi\pi N$ scattering through a complicated analysis plagued with systematic uncertainties, and most experiments (mainly at Berkeley \cite{Pr73} and CERN-Munich \cite{Cern-Munich}) have produced several conflicting data sets, even within the same collaboration, when using different analysis tools. 
As an example, we show in Fig.\ref{fig:00data} the data on $\pi\pi\rightarrow\pi\pi$ scattering phase shifts of the scalar isoscalar wave, were all the $f_0$ states appear. Note the large differences even within data sets coming from the same experiment \cite{Cern-Munich} due to systematic uncertainties. The precise and consistent data sets below 400 MeV all come from $K\rightarrow\pi\pi\ell\nu$ decays \cite{Rosselet:1976pu,Batley:2010zza}, which have almost no systematic uncertainty compared to those from $\pi N\rightarrow\pi\pi N$. Especially relevant for this discussion will be the recent (end of 2010) very precise data from the NA48/2 Collaboration \cite{Batley:2010zza}, since consistency with this data is one of the key requirements for the RPP choice of results for their averages and estimates 
Furthermore, let me emphasize that some of these scalar states and very particularly the $\sigma$, are very wide, or lie close to thresholds, so that the simple Breit-Wigner description, valid for narrow resonances, is not appropriate to describe the data.
Actually, note in Fig.\ref{fig:00data}  that there is no Breit-Wigner shape around 500-600 MeV, corresponding to a $\sigma$ or $f_0(500)$ resonance. In contrast, a Breit-Wigner-like shape over a background phase of about 100 degrees may be seen around 980 MeV, corresponding to the $f_0(980)$, but even that shape is somewhat distorted by the nearby $\bar KK$ threshold.

\begin{figure}[htbp]
  \centering
  \includegraphics[width=1.0\textwidth]{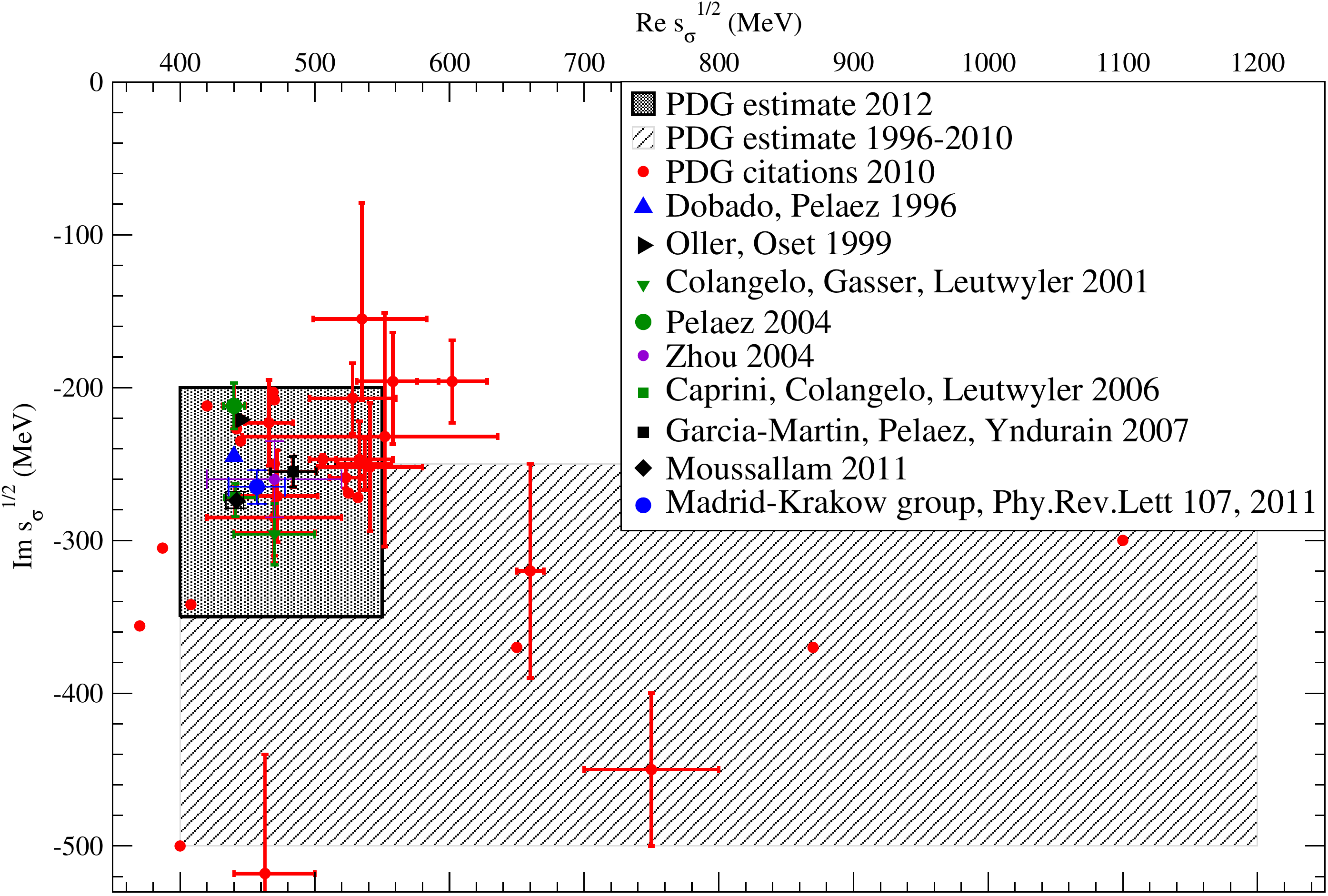}
\vspace*{.1cm}
  \caption{Comparison of the $\sigma$  or $f_0(500)$ resonance poles listed in the RPP
2010 edition, versus that of 2012 \cite{PDG}.
We have highlighted in other colors
those poles obtained from dispersive approaches whether they are recent or old. 
Note the good agreement of the dispersive results, 
all of them concentrated in 
  a small region of the complex plane, 
versus the estimate in the 2010 RPP edition (light gray rectangle)
and the recently revised estimate in 2012 RPP (darker gray rectangle).
}
  \label{fig:poles}
\end{figure}

Since the simple and well known Breit-Wigner resonance approximation is not seen in the experiment, one then has to use the mathematically rigorous definition of a resonance by means of its associated pole in the unphysical (second) Riemann sheet of the complex plane. Still, one keeps the Breit-Wigner notation that relates the pole position $s_R$ with the resonance mass and width as follows: $\sqrt{s_R}\simeq M_R-i \Gamma_R/2$. Consequently, virtually all people working on the scalar mesons refer at some point to this ``pole mass'' and width. This is why the RPP provides the so-called ``t-matrix'' pole since 1996, although, unfortunately, it also provides a Breit-Wigner pole, which, to my view, only leads to confusion, since the $\sigma$ simply cannot be described by a simple Breit-Wigner, as it can be easily seen in Fig.\ref{fig:00data}. Hence, I will restrict my report to the rigorous and sounded ``t-matrix'' pole description, and thus Fig.\ref{fig:poles}shows the position of the $\sigma$ poles in the complex plane. Please note the huge light gray area that corresponds to the uncertainty assigned to the $\sigma$ pole in the RPP until 2010. In addition, the non-red poles correspond to the most recent and some older dispersive approaches that I will comment below. Note that, compared with the huge 2010 huge uncertainty band, the dispersive approaches, and the latest ones in particular, are remarkably consistent in claiming a $\sigma$ pole mass around 450 to 480 MeV. 

At this point it should be clear that, in order to extract the parameters of a pole in the complex plane that lies so far from the real axis as that of $\sigma$, it is not enough to have a good description of the data. The reason is that many functional forms can fit very well some data, but differ widely with each other when extrapolated outside the fitting region. Hence, to look for poles we need the correct analytic extension to the complex plane, or at least a controlled approximation to it. Unfortunately that is not the case in many analyses, so that the poles obtained from a poor analytic extension of an otherwise nice experimental analysis may be artifacts or just plain wrong determinations.

Therefore, a very significant part of the apparent disagreement between different poles in Fig.\ref{fig:00data} is not coming from experimental uncertainties when extracting the data, but from the use of models in the interpretation of that data and the unreliable extrapolation to the complex plane. Actually, the same experiment could provide dramatically different poles, depending on the parametrization or model used to describe the data and its later interpretation in terms of poles and resonances. Maybe the most radical example comes from, on the one hand, the red point sitting in the lowest left corner (at 400-i 500 MeV) and, on the other hand, the one at 1100-i 137 MeV (below the legend), both listed in the RPP tables under the same experimental paper \cite{Amsler:1995bf}. The point around 1100-i300 is also from the same collaboration \cite{Amsler:1995gf}. 

Significant improvement came, over the last decade, from decays of heavier particles into mesons, although there have been no additions in the RPP from the 2010 to the 2012 edition, and I just refer to the RPP \cite{PDG} for older references.  
What is important is that these processes have different systematics than scattering, thus
providing a strong support for a light $\sigma$ below 1 GeV, making the case sufficiently convincing to declare the then called $f_0(600)$ a ``well established'' state in the 2002 RPP edition. Note that, in general, they tend to yield a pole mass somewhat higher than the dispersive approaches, say, between 500 and 550 MeV. Unfortunately
the analysis of these decays is usually performed with much less rigorous and sometimes even inconsistent models, like superimposed Breit-Wigners in isobar models, which violate unitarity, or with K matrices, which should also incorporate information on meson-meson scattering in one way or another. Hence, this information from decays has improved the situation, but its analysis is somewhat still model-dependent.

\subsection{Dispersive approach}

In principle, the rigorous way of determining the poles and residues of the amplitude is by means of dispersion relations. Briefly, in terms of Physics, these relations are a consequence of causality, which 
mathematically allows us to extend analytically the amplitudes to the complex plane of the energy, and then use Cauchy Theorem to relate the amplitude at any value of the complex plane with an integral over the (imaginary part of the) amplitude evaluated in real axis, i.e. the data. Such a relation can be used in several ways. In the physical real axis, it means that the amplitude has to satisfy an integral constraint. Thus, one can check the consistency, within uncertainties, of the data at a given energy against the data that exists in other regions. A stronger possibility is to use the dispersion relations as constraints, by {\it forcing} the amplitude to  satisfy the dispersion relation while fitting the data. Furthermore, one could even use them to obtain values for the amplitude at energies where data do not exist, using existing data in other regions. Finally, once one has an amplitude that satisfies well the dispersion relation and describes the data, it is possible to extend the integral representation to obtain a unique analytic continuation to the complex plane (or at least to a particular region of the complex plane). 
For partial wave amplitudes, one can thus study the complex energy plane and look for poles and their residues, which, as we have explained above, provide the rigorous and observable independent definition for the resonance mass, width and couplings. In principle, by using the integral representation to perform the analytic continuation, the particular model or functional form chosen to parametrize the data becomes irrelevant and the spectroscopic results are parametrization and model-independent.

Typically, dispersion relations for relativistic scattering are formulated in terms of the Mandelstam variable $s$, by getting rid of the $t$ dependence either by fixing or integrating it.
Thus, on the one hand, when $t$ is fixed we talk about ``fixed-$t$ dispersion relations''.
Once $t$ is fixed, one can usually choose combinations of amplitudes which are symmetric or antisymmetric under $s\leftrightarrow u$ exchange and exploit these symmetries to write a dispersion relation, which involves integrals only over the physical region.
Of special importance among this kind of dispersion relations is the case when one fixes $t=0$, known as ``Forward Dispersion Relations'' (FDRs). They are very relevant because, due to the optical theorem, the imaginary part of the forward amplitude is proportional to the total cross section, and data on total cross sections are generically more abundant and of better quality than on amplitudes for arbitrary values of $s$ and $t$. Thus, the forward dispersive integrals are usually calculable and very reliable for FDRs.

On the other hand, one could integrate $t$ by projecting the amplitude into partial waves $f(s)$, for which the dispersion relation is then written. The advantage of these partial wave dispersion relations is that their poles in the second Riemann sheet are easily identified as resonant states with the quantum numbers of the partial wave. Therefore they are very interesting for spectroscopy.
However, due to crossing symmetry, partial waves have a ``left cut'' in the unphysical $s$ region, which also contributes to the dispersion relation. If the region of interest lies very far from this cut, it could be neglected or approximated, but when closer, or if one wants to reach a good level of precision, it could be numerically relevant and has to be taken into account. This is the case of several of the resonances of interest for us, namely, the $\sigma$ and the $K_0^*(800)$ which are very deep in the complex plane and relatively close to threshold and to the left cut.
But  including correctly the left cut is somewhat complicated because that unphysical energy region may correspond, due to crossing symmetry, to different processes arising from crossed channels in other kinematic regions and other partial waves. In addition one may not have data for these other processes. Dealing rigorously with the left cut usually involves an infinite set of coupled integral equations. These were formulated long ago for $\pi\pi\rightarrow\pi\pi$ scattering, known as Roy equations \cite{Roy:1971tc}, and have received considerable attention over the last decade \cite{ACGL,CGL,Kaminski:2002pe,PY05,Caprini:2005zr,GKPY11,GarciaMartin:2011jx,Moussallam:2011zg}.  These Roy equations were already widely used from the 70's, but their accuracy was limited by the lack of precise and reliable data at threshold. This caveat can be circumvented in two ways, either by the use of Chiral Perturbation Theory predictions at low energy, as in \cite{CGL},
or, if one wants to avoid the use of further theoretical input as in \cite{GKPY11}, by the use of the very recent and precise data from NA48/2 in 2010 \cite{Batley:2010zza}. The former made it possible to make a very precise determination of the $\sigma$ pole, using ChPT predictions, also showing that Roy Equations provided a consistent analytic extension in the complex plane that reaches the area where the sigma pole is found \cite{Caprini:2005zr}. The latter, which is nothing more than a dispersive data analysis has been followed by our Madrid-Krakow Collaboration \cite{GKPY11,GarciaMartin:2011jx} and required the derivation of another set of Roy-like Equations, called GKPY Equations, but with one subtraction (further energy suppression in the dispersive integrals) instead of two as in the original Roy derivation \cite{Roy:1971tc}. 

Fortunately, the dispersive formalism is very powerful and relatively simple for $\pi\pi\rightarrow\pi\pi$, where the $\sigma$ appears in the scalar-isoscalar partial wave,  and the latest dispersive analyses have provided strong support for the existence of such an state and also have provided the best determinations of the $\sigma$ properties. Thus, in Fig.\ref{fig:poles} we have highlighted in colors other than red, determinations of the $\sigma$ pole based on dispersion relations, including the latest ones of 2011 selected in the 2012 RPP edition. It can be noticed that, within the community working on light scalars, the existence of a light scalar with a pole 
around 500 MeV has been rather well known for quite a few years. The differences or uncertainties are on the range of a few tens of MeV for the mass, not several hundreds. There was a caveat on the size of the ``left cut'' contribution, that was efficiently calculated in \cite{Caprini:2005zr}, obtaining a very precise result, which has been confirmed by our group \cite{GarciaMartin:2011jx}, but without using any ChPT input and the low energy NA48/2 data instead.
This, together with new and precise data close to $\pi\pi$ threshold from NA48/2 in 2010 \cite{Batley:2010zza}, has thus triggered a major revision in the 2012 RPP edition reducing the uncertainty in the $\sigma$ mas by a factor of more than five, leaving it from 400 to 550 MeV,
 and by almost a factor of two for the width, which is now quoted to be between 400 and 700 MeV.
This ``estimate'' takes into account, not only the very rigorous dispersive analyses, but 
other results from models which are required to be at least consistent with the accurate $K\rightarrow\pi\pi\ell\nu$ decay results from NA48/2 \cite{Batley:2010zza} and \cite{Pislak:2001bf}, as well as  experimental values from other processes like heavy meson decays, which, as commented above, yield somewhat larger masses, and do not use dispersive techniques to extract the pole rigorously, but just some models. 
The new RPP uncertainty band is shown in Fig.\ref{fig:poles} as the smaller and darker rectangle.
Accordingly, even the name of the particle has been changed to $f_0(500)$.
The RPP also provides Breit-Wigner parameters, but I would rather not comment on these for the reasons explained above.

To my view, these RPP criteria are still rather conservative, and in the case of the $\sigma$ I would only rely on the 
very rigorous extractions of the poles, which take care of all analytic constraints. But, admittedly, this major revision constitutes a very considerable and long awaited improvement upon the previous confusing situation. 

\begin{figure}[htbp]
  \centering
  \includegraphics[width=1.0\textwidth]{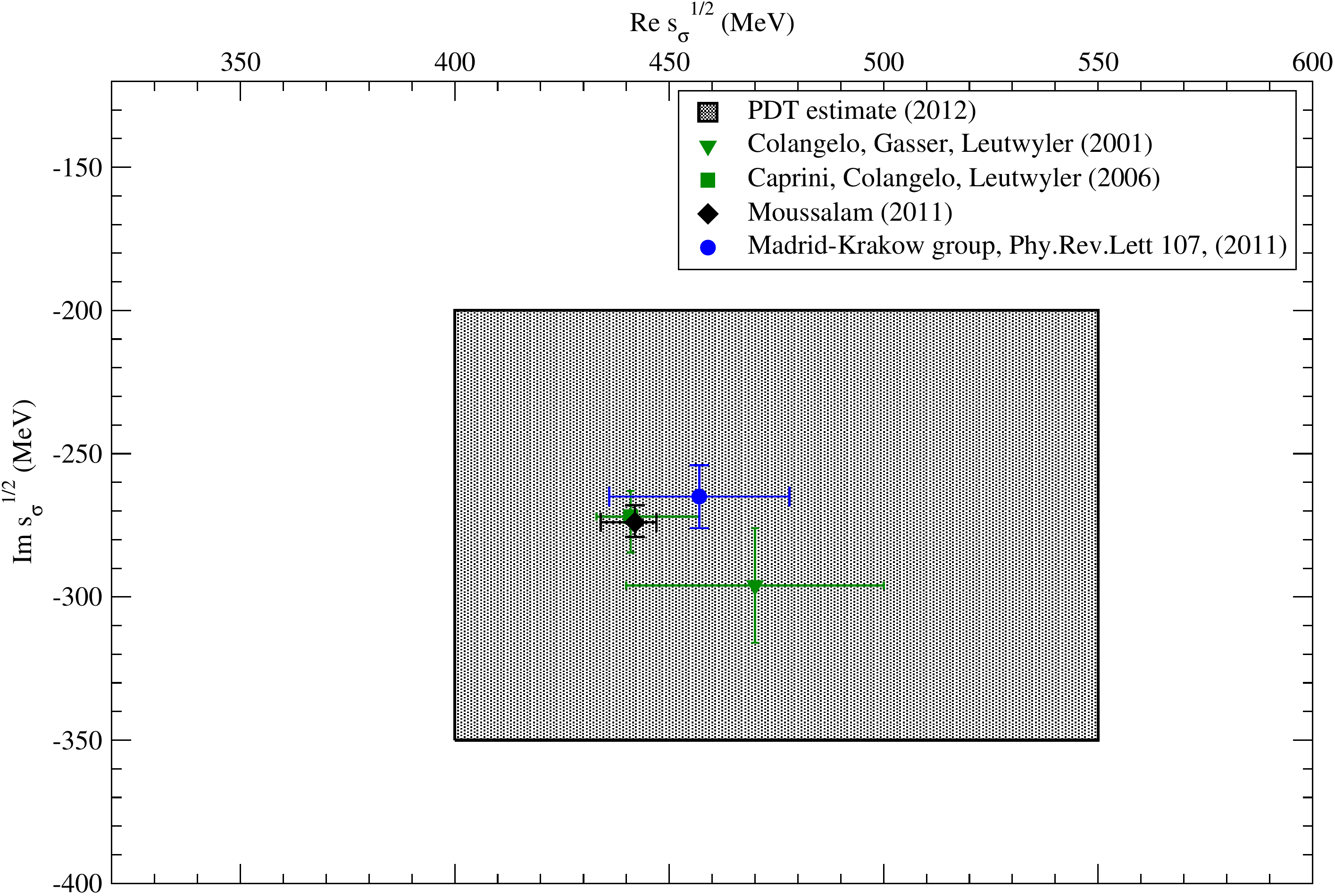}
\vspace*{.1cm}
   \caption{The four ``most advanced dispersive analyses'' \cite{CGL,Caprini:2005zr,GarciaMartin:2011jx,Moussallam:2011zg} according to the ``Note on light scalars'' of the 2012 RPP \cite{PDG}, which lead
 to their ``more radical ...'' average $M-i\Gamma/2\simeq\sqrt{s_\sigma}=(446\pm6)-(276\pm5)\,$ MeV. Note their consistency and that they provide a much more precise determination that the present conservative estimate in the RPP, $M=450$ to 550 MeV, $\Gamma= 400$ to 700 MeV, which we show as a gray box. Note that this gray box here corresponds to the small and darker box in 
 Fig.1
and is already much smaller than the uncertainty in the previous 2010 editions.
 }
  \label{fig:disppoles}
\end{figure}

The significance of the use of dispersion relations in this RPP revision can be gauged by noting that, actually, the RPP is well aware of the caveats that I have just pointed out above on the extraction of poles which are so deep in the complex plane, and thus the RPP `Note on light scalars'' suggests that 
one could ``take the more radical point of view and just
average the most advanced dispersive analyses''. The RPP choice corresponds to references \cite{CGL,Caprini:2005zr,GarciaMartin:2011jx,Moussallam:2011zg} here, and are shown in Fig.\ref{fig:disppoles}. 
The 2012 RPP  
average yields a pole
at $M-i\Gamma/2\simeq\sqrt{s_\sigma}=(446\pm6)-(276\pm5)\,$ MeV.

\begin{figure}[htbp]

\hspace*{-.6cm}
  \includegraphics[width=.54\textwidth]{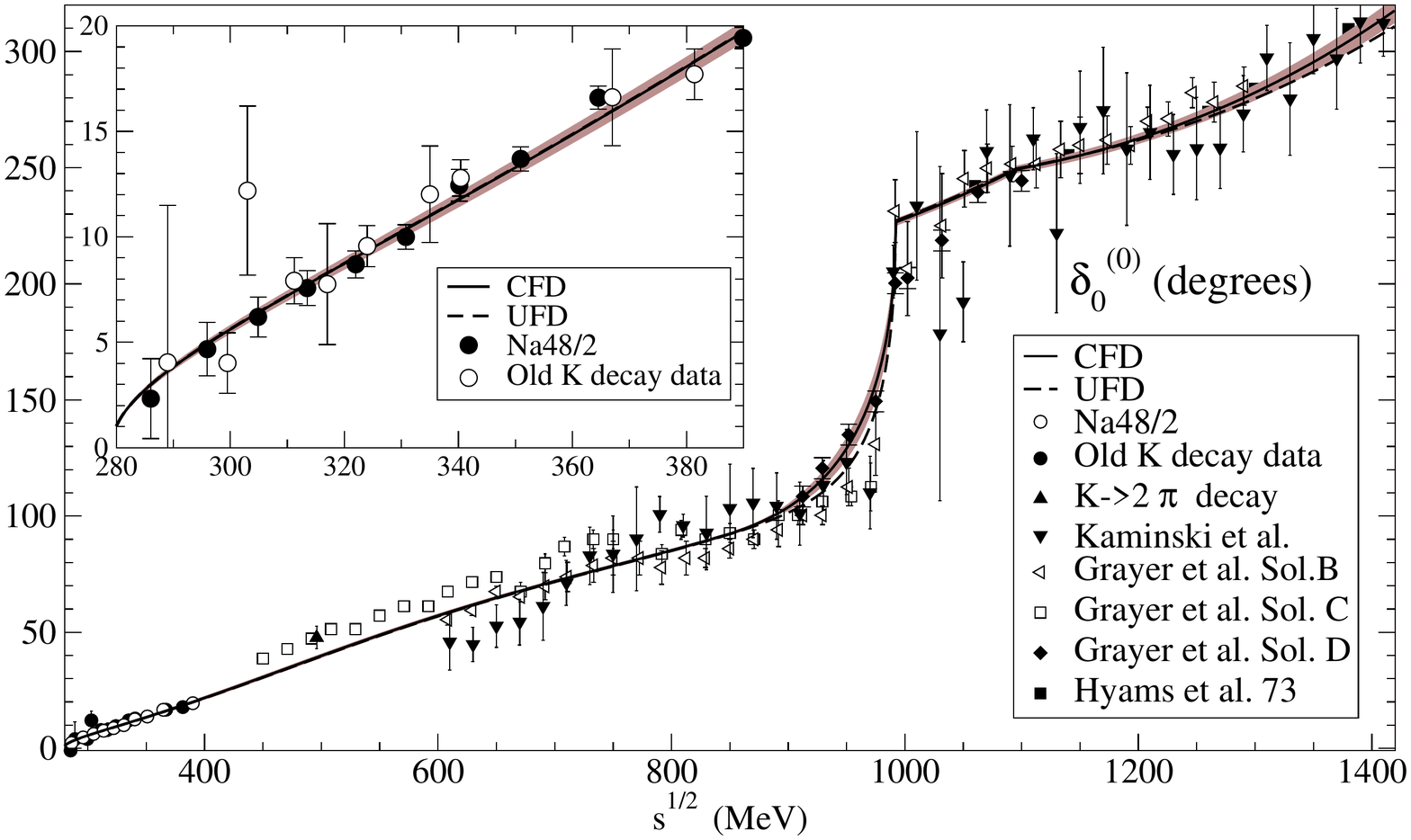}
\hspace*{-.7cm}
  \includegraphics[width=.53\textwidth]{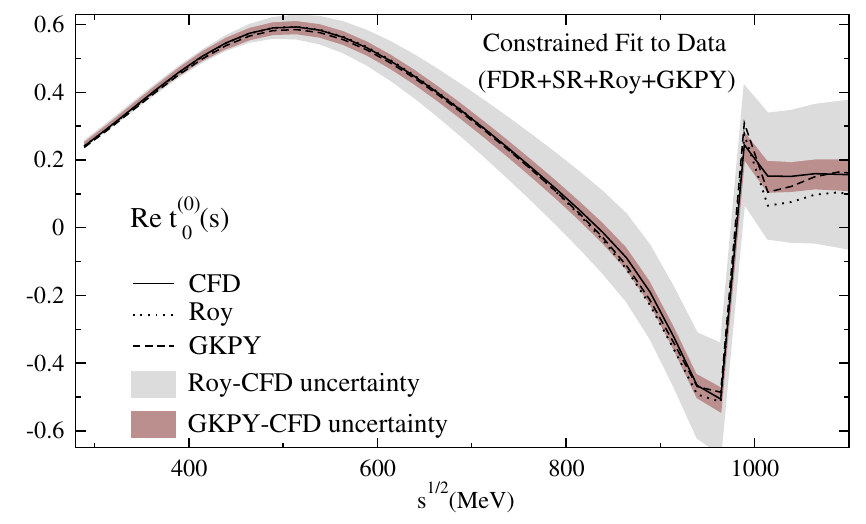}
  \caption{ Left panel: scalar-isoscalar $\pi\pi$ scattering phase from UFD and CFD parametrizations from \cite{GKPY11}. The dark bands cover the uncertainties.
Right panel: Fulfillment of the Roy and GKPY equations for the scalar-isoscalar wave. 
The CFD parametrization is the input to both the Roy and GKPY eqs.,
and is in remarkable agreement with their output.
Note how the uncertainty in the Roy eq. is much larger 
than that of the GKPY eq. above roughly 500 MeV.
}
  \label{fig:UFDCFD}
\end{figure}

In order to illustrate how these dispersive techniques work, let me now describe, as an example, the procedure followed by our Madrid-Krakow Collaboration \cite{GKPY11,GarciaMartin:2011jx} to obtain these fits. We use, as a starting point, a set of Unconstrained Fits to Data (UFD) which was shown to be not too inconsistent with forward dispersion relations. In a second step, one modifies slightly the parameters of these fits to satisfy the dispersion relations, without spoiling the description of the data. This new set is called ``Constrained Fits to Data'' (CFD). Both the CFD and UFD parametrizations for the scalar isoscalar $\pi\pi$ scattering phase shift are shown on the left panel of Fig.\ref{fig:UFDCFD}, where the very low energy region, which drives the size of the uncertainties is show in an inner box. The only sizable differences between the UFD and CFD appear, in the 1 GeV region and above, but they are small enough so that both provide a very good description of the data. 
In the right panel of Fig.\ref{fig:UFDCFD} we then show how well the CFD set satisfies, for instance, the Roy and GKPY equations for the real part of the scalar-isoscalar wave $t_0^{(0)}$. The continuous line is the input obtained from the CFD parametrization, whereas the dotted and dashed lines are the output of the dispersive Roy and GKPY representations, respectively. If these equations were satisfied exactly, they would coincide, but we just ask them to overlap within the uncertainties. Note that the once-subtracted GKPY equations are more precise in the resonance region, say above 500 MeV, whereas Roy equations are more accurate below that energy, given the same input.
The final step, once a data description consistent with a whole set of dispersion relations, unitarity and symmetry constraints, etc... has been obtained, is to use the dispersion relation to continue the amplitude into the complex plane and look for poles in the unphysical sheets, which are associated to resonances. In principle, the use of a dispersion relation to perform the analytic continuation implies that the resulting pole is model independent, and, in particular it cannot result from an artifact of the functional form used to fit the data. This is actually what was done in \cite{GarciaMartin:2011jx} with the previous CFD parametrization used as input of the GKPY or Roy equations, which are then used to continue the amplitude to the complex plane, with the final result:
\begin{eqnarray}
\sqrt{s_\sigma}&=&(457^{+14}_{-13})-i(279^{+11}_{-7})\; {\rm MeV}\qquad {\rm(from \, GKPY \,eqs.)} \label{myresult},\\
\sqrt{s_\sigma}&=&(445\pm25)-i(278^{+22}_{-18})\; {\rm MeV} \quad \,\,{\rm(from \,Roy \,eqs.)}.
\end{eqnarray}

Our two results just above are one of the five new entries in the 2012 RPP edition. The other new entries are two results from an ``analytic K-matrix model'' in \cite{Mennessier:2010xg}:  $(452\pm13)-i(259\pm16)\, {\rm MeV}$ and $(448\pm43)-i(266\pm43)\, {\rm MeV}$, depending on what data sets and different variants of the K-matrix model are averaged. Finally, the other new result in the 2012RPP is $(442^{+5}_{-8})-i(274^{+6}_{-5})\, {\rm MeV}$ from \cite{Moussallam:2011zg}. The latter is also based on Roy equations, which uses as input for other waves and higher energies
the Roy equations output of \cite{CGL} and is therefore very consistent with the older result in \cite{Caprini:2005zr}: $\sqrt{s_\sigma}=(441^{+16}_{-8})-i(272^{+9}_{-12.5})\, {\rm MeV}$, which used ChPT input, as well as with that even older in \cite{ACGL}: $(452\pm13)-i(259\pm16)\, {\rm MeV}$. These last three results, based on Roy equations, together with those two in Eq.(\ref{myresult}) above, are precisely those ones considered by the RPP as the ``most advances dispersive analyses'', shown in Fig.\ref{fig:disppoles} here.

Once more, I would like to remark that the major RPP revision of the  $\sigma$
has only acknowledged a result---the existence of a very wide resonance around 500 MeV, or rigorously, a pole deep in the complex plane of the scalar isoscalar amplitude---which was a well known fact in the community. The dispersive analysis have just helped providing further rigor and precision to many other studies that implied the existence of such object, as can be seen in the Fig.\ref{fig:poles}. 

Finally, I would like to comment on a question that was asked at the end of this talk, concerning the nature of the $\sigma$ in connection to its $N_c$ behavior. Of course, once the existence of the $\sigma$ as a light and very wide resonance around 500 MeV was settled, 
the debate focused on its nature and interpretation from QCD.
Actually, the model independent dispersive techniques which are so appropriate for an accurate determination of the pole, usually tell us little about the composition of the states, whether they form multiplets, etc... Hence, answering the question: Yes, in principle, the $N_c$ behavior of the resonance can tell us about the composition of the state, but for that we need theory input, or a model, to implement the $N_c$ behavior according to the QCD rules. And this is precisely what we have been trying to avoid with the use of model independent dispersion theory.  Then, for the purposes of understanding the nature and spectroscopic classification, instead of the most rigorous dispersive approaches, one could also use models that have at least the most relevant analytic structure, impose further constraints in the form of sum rules, and make sure that the resonances claimed lie within the applicability of the approach. Some models, based or those using Effective Field Theories and simplified dispersion relations, as for instance the N/D method or some unitarized models,  can be very useful to obtain resonance poles and parameters in cases with coupled channels, at least in those channels were reliable data exist (see, for instance \cite{others,otherssigmakappa}), and also to study their behavior and interpretation, as long as they are able to reproduce, within a good approximation, the same poles found above with dispersion theory. In particular, the $N_c$ behavior of the $\sigma$ has been studied \cite{largen} using dispersion relations for the inverse of the amplitude and calculating the subtraction constants with the ChPT expansion, whose $N_c$ behavior is known, and the results indicate that predominantly, it might not be an ordinary $\bar qq$ meson, although it might have some small $\bar qq$ component, which originates above 1 GeV \cite{largen}. For further details on this issue, I refer to the original works on $N_c$ and the $\sigma$ \cite{largen} as well as to the ``Note on light scalars'' of the RPP \cite{PDG}, and references therein.

\section{The $f_0(980)$}

The determination of the $f_0(980)$ parameters has been much less controversial (although not the interpretation  of its nature). In the 1990 RPP, for instance it was listed as
the $f_0(975)$, with an estimated mass of $975.6\pm3.1$~ MeV and a width of $33.6\pm5.6$, whereas in the 1994 edition it was quoted as the $f_0(980)$ with an estimated mass of $980\pm10$ MeV and a width from 40 to 100 MeV. The latter estimates we kept until the 2012 edition. The reason for this stability, when compared with the sigma, is already obvious from Fig.\ref{fig:00data}, although $\pi\pi\rightarrow\pi\pi$ is not the best place to look for this resonance:
there is a narrow resonance shape near 980 MeV over a $\sim 100$ degree background phase.
This background is supposedly due to the long tails of the wide $\sigma$ and the $f_0(1370)$.
Note also that this resonance is very close to the $K^+ K^-$  and $\bar K^0 K^0$ thresholds, which differ by about 8 MeV, due to isospin breaking, which is usually neglected in most approaches, and is an effect that contributes significantly to the uncertainty in the $f_0(980)$ parameters, particularly, the width.

\begin{figure}[htbp]
  \centering
  \includegraphics[width=.43\textwidth]{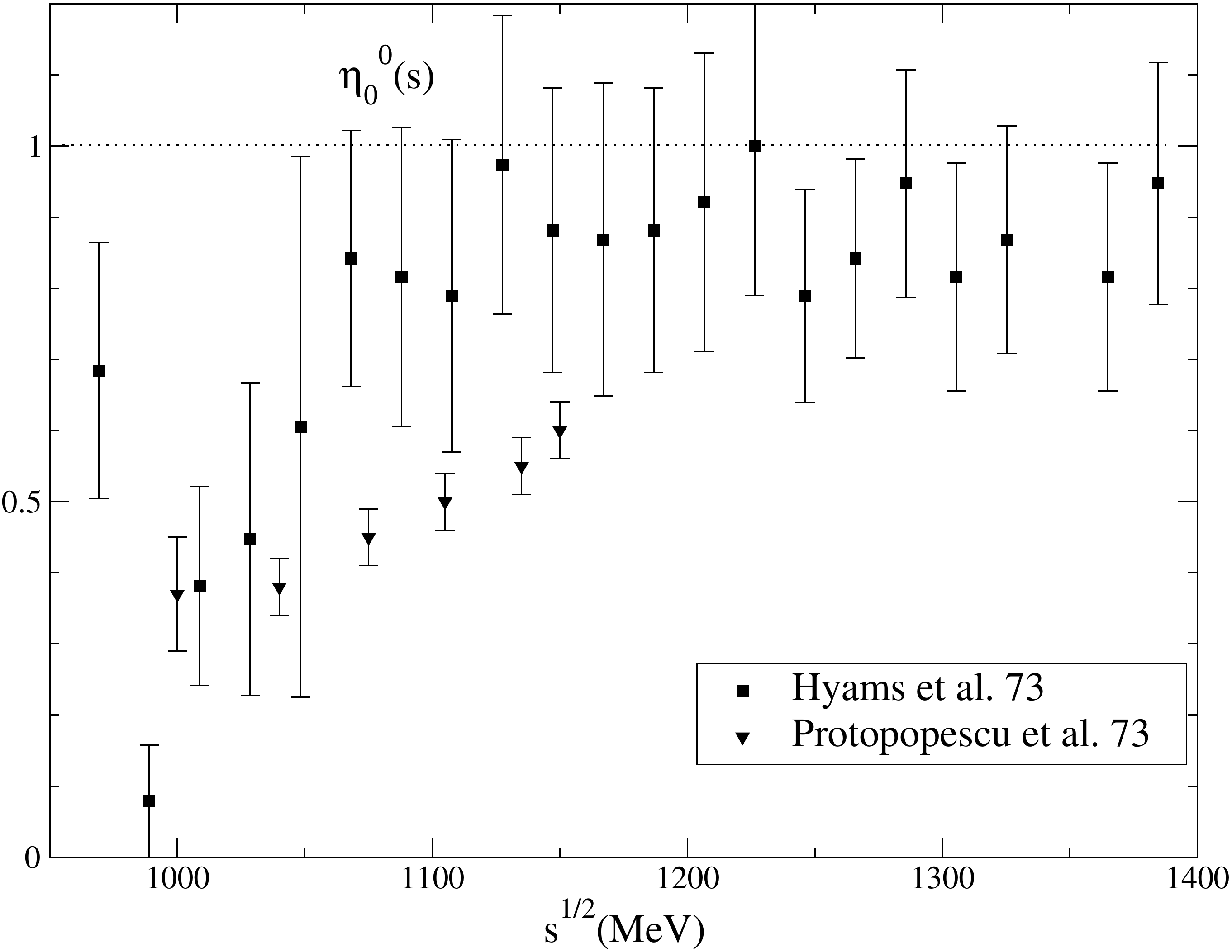}
  \includegraphics[width=.43\textwidth]{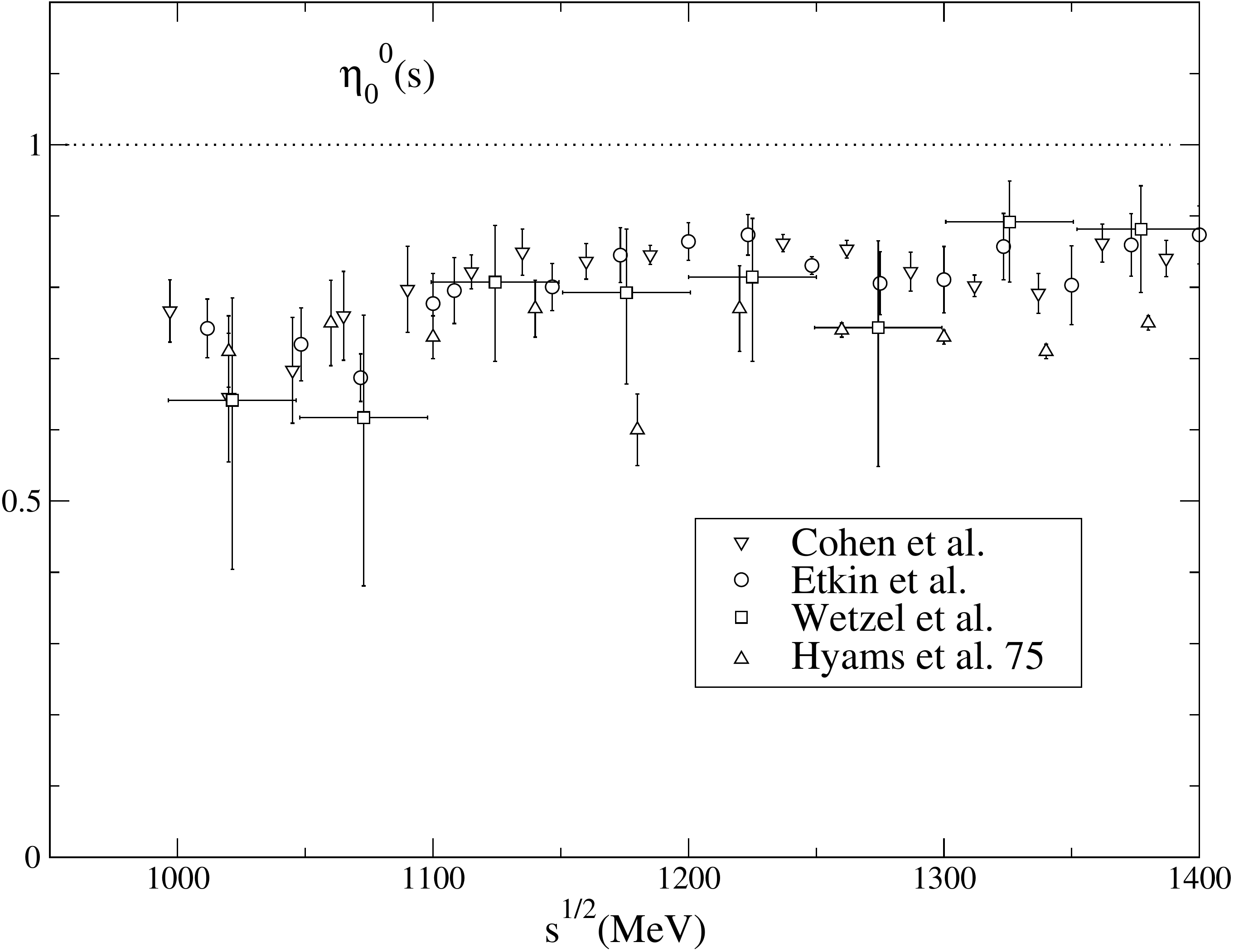}
  \includegraphics[width=.43\textwidth]{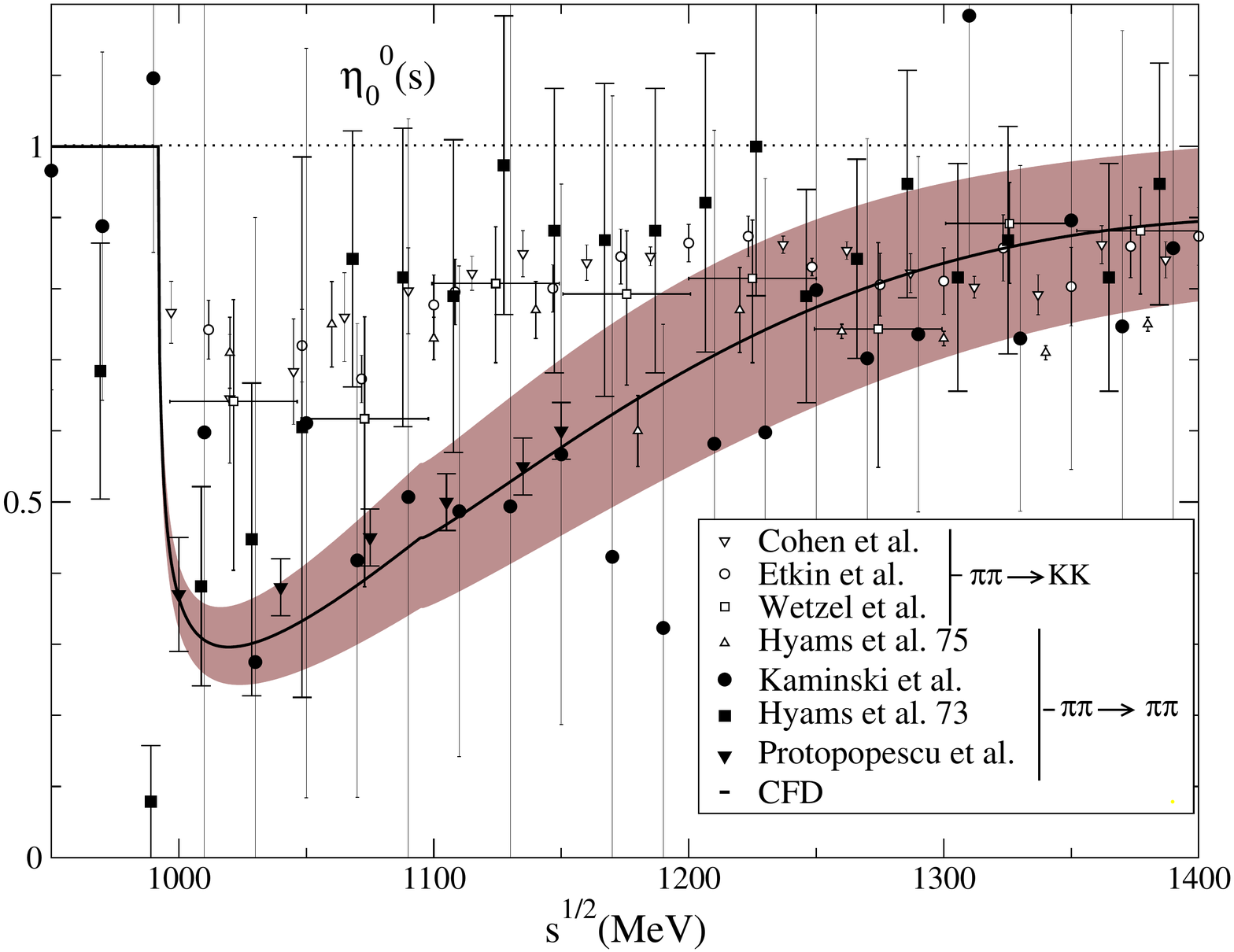}
\vspace*{.1cm}
  \caption{ ``Dip'' and ``no-dip'' scenarios for the $\pi\pi$ scattering inelasticity (Left and right upper panels, respectively), as well as the solution from the Constrained Fits to Data, satisfying the dispersive representation in \cite{GKPY11} 
}
  \label{fig:S0inel}
\end{figure}

Nevertheless, after almost two decades of keeping the same estimate, the 2012 RPP edition has made a small update and now quotes a mass of $990\pm20$ MeV, keeping the same large width uncertainty. As pointed out in the 2012 ``Note on light scalars'', this small shift on the mass and the doubling of the uncertainty was updated to accommodate the recent dispersive analysis 
by our group \cite{GKPY11,GarciaMartin:2011jx}, described in the previous section.
We obtain an $f_0(980)$ pole (in the second Riemann sheet) at: $\sqrt{s_{f_0(980)}}=(996\pm7)-i(25^{+10}_{-6})$ MeV if we use GKPY equations and $\sqrt{s_{f_0(980)}}=(1003^{+5}_{-27})-i(21^{+10}_{-8})$ MeV
from Roy equations. The relevance for the $f_0(980)$ parameters of our dispersive study is that it has settled a longstanding conflict between two kind of solutions for the elasticity parameter in $\pi\pi$ scattering, which is almost zero up to $\bar KK$ threshold, and is later dominated by $\pi\pi\rightarrow\bar KK$. The two conflicting solutions are known as ``dip-scenario'', shown in the upper left panel of Fig.\ref{fig:S0inel}, and the ``non-dip scenario'' shown in the upper right panel of the same figure. In \cite{GKPY11} it was shown that the dip scenario satisfies well the GKPY dispersion relations as seen in the lower panel of Fig.\ref{fig:S0inel}, whereas it is not possible to accommodate the non-dip scenario, because even a relatively poor fit would spoil the simultaneous description of the phase. That the ``no-dip'' scenario is disfavored has later been confirmed in \cite{Moussallam:2011zg} using Roy equations, also obtaining a pole at: $\sqrt{s_{f_0(980)}}=(996^{+4}_{-14})-i(24^{+11}_{-3})$ MeV, in remarkable agreement with our group's findings. Actually, these three dispersive values together with the one from the ``analytic K-matrix'' approach in \cite{Mennessier:2010xg}, which yields $\sqrt{s_{f_0(980)}}=(981\pm43)-i(18\pm11)$ MeV, are the only new additions to the $f_0(980)$ RPP tables in the 2012 edition.

\section{The $K_0^*(800)$}
The $K_0^*(800)$, also traditionally known as $\kappa$ resonance, is still ``omitted from the summary table'' because it is listed in the RPP tables under ``needs confirmation''.
To my view, this is rather surprising, because virtually all descriptions of the data 
with correct implementations of unitarity, chiral symmetry, and some minimal analyticity requirements \cite{otherssigmakappa,otherskappa}, find a pole around
650 to 770 MeV with a 550 MeV width or larger. For some time it was suggested that the state could be more massive, but the possibility of a $\kappa$ lighter than the $K^*(1430)$ but above 825 MeV was discarded in \cite{Cherry:2000ut} using methods based on analyticity.
Actually, in the detailed tables it is listed 
with a mass of $685\pm29$ MeV, and a width of $547\pm24$ (less uncertainty than the $f_0(500)$, although, in contrast, the $f_0(500)$ is already considered a''well established'' state).
In addition, it has also been found in decays of heavier mesons. It is true the latter often suffer from the same caveats that complicated the $\sigma$ analysis, namely, the strong model dependency of the models used to extract the resonance parameters, sometimes in terms of Breit-Wigner amplitudes, isobars models, etc...
However, this was already the case for the $\sigma$, and although these analyses may be questionable in terms of precision, they clearly agree on the need for a pole in that region.

Furthermore, a version of the Roy dispersion relations exists in the $K \pi\rightarrow K\pi$ scattering. These are called Roy-Steiner equations \cite{Steiner:1971ms} and are somewhat more 
complicated due to the fact that the scattering particles are different, and thus a different process, namely the crossed channel reaction $\pi\pi\rightarrow \bar KK$ is also needed as input.
However, this rigorous analysis has been performed \cite{Buettiker:2003pp} and even though
a Breit-Wigner phase motion is not seen the scattering data (as we have seen it also happens with the $f_0(500)$) the dispersive approach requires the existence of a pole at $(658\pm13)-i(278.5\pm12)$ MeV \cite{DescotesGenon:2006uk}.

Part of the confusion may arise from the use of incorrect Breit-Wigner parametrizations for such a wide state that requires a pole deep in the complex plane, which  in addition is near the $K\pi$ threshold and very close to the left cut (and the circular cut that appears due to the different masses of the scattering particles). The shape seen in the data in the real axis is a combination of all these effects plus the chiral symmetry requirement of a so-called Adler zero near threshold. Contrary to the case of the $\sigma$, the RPP makes no distinction of Breit-Wigner parameters and ``t-matrix" pole parameters. Thus, the new additions in the $K_0^*(800)$ come from the Breit-Wigner parameters of a study of $J/\Psi$ decays at BES2, which obtain for the mass $M=849\pm77^{+18}_{-14}$ MeV or $M=826\pm49^{+49}_{-34}$ MeV and for the width $\Gamma=512\pm80^{+ 92}_{-44}$ MeV or $\Gamma=449\pm156^{+ 144}_{-81}$ MeV, depending on whether there is a charged \cite{Ablikim:2010kd} or a neutral \cite{Ablikim:2010ab} kaon in the decay products, respectively. 
The central values for these masses may seem somewhat larger than the range of 650 to 750 MeV
I gave for the pole above, but let me remark again that these are Breit-Wigner parameters
and that, although not listed in the RPP, the BES2 Collaboration in \cite{Ablikim:2010ab} also provides a pole position
$764\pm^{+71}_{-54}-i(306\pm149^{+143}_{-85})$ MeV which is more consistent with results form scattering.

In conclusion, given that: a) this state appears as a wide pole in scattering, not only in well sounded models but also in rigorous dispersive analyses---as it happened with the $\sigma$.
b) A quite consistent pole is also seen in heavy meson decays, which have very different systematic uncertainties from scattering, as it also happened with the $\sigma$. Then, I definitely support that it should be considered another ``well established'' light meson state and treated on a similar footing as the $f_0(500)$ resonance.

\section{The $a_0(980)$}

As with the $f_0(980)$, the existence of this state is not controversial, although the uncertainties are somewhat larger. Its parameters have been relatively stable throughout the RPP editions. Indeed, there have been no relevant developments for the $a_0(980)$ since the last edition of this Conference, and, accordingly, no additions to the $a_0(980)$ tables in the RPP 2012 edition.

\section*{Acknowledgments}
I would like to thank the organizers of the Conference for inviting me to present this review talk and for creating such a nice atmosphere for discussing Physics, as well as for their kind understanding and help with my last minute changes in my participation.
This work is supported in part
 by the Spanish Research contracts
FPA2007-29115-E/, FPA2011-27853-C02-02 and  the EU FP7 HadronPhysics3 project.

\end{document}